%% file: babar-pub-14008.tex
\documentclass[aps,prd,floatfix,twocolumn,preprintnumbers,groupedaddress,superscriptaddress,altaffilletter,nobibnotes,amsmath,amssymb,amsfonts,showpacs]{revtex4-1}
\usepackage{color}
\usepackage{graphicx}
\usepackage{dcolumn}
\usepackage{bm}
\usepackage{feynmf}
\usepackage{tikz}
\RequirePackage{snapshot}

\input{babarsym}

\def\fQ2         {\ensuremath{\mathcal{F}_{\gaga\to\piz}\left(Q^2\right)}\xspace}
\def\fQQ         {\ensuremath{\mathcal{F}_{\gaga\to\piz}\left(Q^2\right)}\xspace}
\def\fQQ         {\ensuremath{\mathcal{F}_{\piz}\left(Q^2\right)}\xspace}
\def\impS        {\ensuremath{\phi_S}\xspace}
\def\impP        {\ensuremath{\phi_P}\xspace}
\def\impHCP      {\ensuremath{\piz_{\textrm{HC}}}\xspace}
\def\impost      {\ensuremath{\phi}\xspace}
\def\Eon         {\ensuremath{\sqrt{s}=10.58\gev\xspace}}
\def\Eoff        {\ensuremath{\sqrt{s}=10.54\gev\xspace}}
\def\stat        {\text{~\footnotesize(stat.)}}
\def\syst        {\text{~\footnotesize(syst.)}}
\def\coloronline {\small(color online) \xspace}

\begin{document}


\preprint{%
\begin{minipage}[l]{\textwidth}
{SLAC-PUB-16139}\\
{BABAR-PUB-14/008}
\end{minipage}
}

\title{Search for new \piz-like particles produced in association with a \mtau-lepton pair}

\input{authors_aug2014_bad2598.tex}

\begin{abstract}
  We report on a search in \epem annihilations for new \piz-like
  particles produced in association with a \mtau-lepton pair. These
  objects, with a similar mass and similar decay modes to \piz mesons,
  could provide an explanation for the non-asymptotic behavior of the
  pion-photon transition form factor observed by the \babar\
  Collaboration.  No significant signal is observed, and limits on the
  production cross section at the level of 73\fb or 370\fb, depending
  on the model parameters, are determined at 90\% confidence level.
  These upper limits lie below the cross section values needed to
  explain the \babar\  form factor data.
\end{abstract}

\pacs{14.40.Rt, 14.60.Fg}

\maketitle


\section{Introduction}
\label{sec:Introduction}

The measurement of the pion-photon transition form factor \fQQ
reported by the \babar\ Collaboration~\cite{Aubert2009} has given rise
to much
discussion~\cite{Bakulev2012,Dorokhov2010,Lucha2012,Noguera2010}. 
The result does not exhibit convergence towards the Brodsky-Lepage
limit of $185\mev/Q^2$~\cite{Brodsky1983} even for large values of
the squared momentum transfer, viz., $Q^2>15\gev^2$, where the data are
expected to be well described by perturbative QCD.  Results from the
Belle Collaboration~\cite{Uehara2012} show better agreement with the
perturbative predictions but are consistent with the \babar\ data within
the uncertainties.

A recent suggestion~\cite{McKeen2012} proposes that the observed lack
of asymptotic behavior might be due to the production of new particles or
states, tentatively named ``pion impostors'' and generically denoted
\impost~\footnote{\impost represents a new object not related to the
  $\phi(1020)$}.  Two classes of models are considered. In the first,
scalar \impS or pseudoscalar \impP particles are introduced with a
mass within 10\mevcc of the \piz\ mass, and with similar decay
modes to the \piz, such that they thereby contribute to the \fQQ measurement.
In the second, a new light pseudoscalar state mixes with the \piz to
produce a so-called ``hardcore pion'' \impHCP.  The \impP and \impHCP
have similar experimental signatures and the related processes only
differ in their production rates. These models predict large coupling
strengths between the new objects and the \mtau lepton, comparable to
the strength of the strong force, leading to an observable increase
of \fQQ through virtual loops with \mtau leptons.  The couplings
of the new particles to heavy quarks and other Standard Model (SM)
particles are constrained by experimental data to be an order of
magnitude or more smaller~\cite{McKeen2012}.

The largeness of the predicted couplings of the pion impostors to the
\mtau lepton, and the absence of corresponding experimental
constraints, motivate a search for pion impostors radiated from \mtau
leptons in $\epem \to \tautau \phi$, $\phi\to\gaga$ interactions.
This process is particularly compelling because the rate of such
events must be considerable in order to explain the \babar\ \fQQ
data, making it potentially observable.  The production cross sections
required to describe the \babar\ measurements are listed in
Table~\ref{tab:prodxsec}. The corresponding results for the
combined \babar\ and Belle data are also given.  Based on the cross
sections derived from the \babar\ data alone, on the order of $10^5$
events are expected in the \babar\ data sample.

\begin{table}[ht!]
  \caption{Production cross sections of \epem\to\tautau\impHCP,
    \tautau\impP, and \tautau\impS at \Eon\ needed
    to accommodate the pion-photon transition form
    factor reported by \babar, as well as the combination of \babar\ and
    Belle measurements. Confidence intervals at 95\% confidence level
    are provided in brackets.}
\begin{ruledtabular}
\begin{tabular}{lrlrl}
\textrm{Model}   & \multicolumn{2}{c}{$\sigma$(pb)}   & \multicolumn{2}{c}{$\sigma$(pb)} \\
                 & \multicolumn{2}{c}{\textrm{\babar~\cite{Aubert2009}}} &
                   \multicolumn{2}{c}{\textrm{\babar\ + Belle~\cite{Uehara2012}}} \\
\colrule 
\impHCP & 0.62   &[0.25 -- 0.84] & 0.44    & [0.15 -- 0.59]   \rule{0pt}{12pt}\\
\impP   & 4.8    &  [2.5 -- 6.9] & 3.4     & [  2.5 -- 5.1]  \rule{0pt}{10pt}\\
\impS   & 130    &   [70 -- 180] & 90      & [   50   -- 140]  \rule{0pt}{10pt}\\
\end{tabular}
\label{tab:prodxsec}
\end{ruledtabular}
\end{table}

\begin{figure}[t!]
  \centering
  \includegraphics[scale=1]{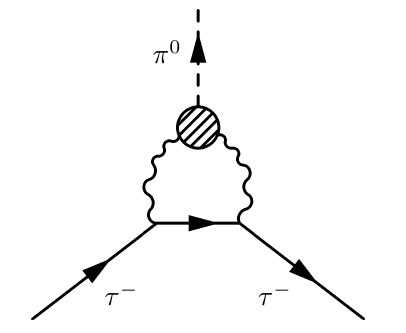}
  \caption{Diagram of the leading order SM process for \piz radiation from a \mtau lepton.}
  \label{fig:feyn_SM_tau_to_pi0}
\end{figure}

The SM production of genuine \piz meson in association
with a \mtau-lepton pair is expected to be highly suppressed. To lowest order, the
SM process in which  a \piz is radiated from a \mtau lepton is depicted in
Fig.~\ref{fig:feyn_SM_tau_to_pi0}.
The matrix element involves the pseudoscalar to two-photon transition
amplitude as well as a suppression factor arising from the two-photon
loop and the \mtau-lepton propagator.  The matrix element for this
diagram~\cite{Bergstroem1982,Bergstroem1983} yields an effective
coupling between the $\piz$ and the \mtau lepton of the form
\begin{equation}
g_{\tau\tau}^\text{e.m.} = - \frac{1}{\sqrt{2}}\frac{m_\tau}{f_\pi}\Bigl(\frac{\alpha}{\pi}\Bigr)^2 R \text{,}
\end{equation}
where $m_{\tau}$ is the mass of the \mtau lepton,
$f_\pi\simeq0.130\gev$ is the pion decay constant, and $\alpha$ is the
fine-structure constant. The factor $R$ is a dimensionless complex
amplitude that is a function of the pion form factor
$\mathcal{F}_{\piz}\left(k^2,(p_{\piz}-k)^2\right)$, integrated over
the virtual photon four-momentum $k$, and of the mass ratio $m_\tau /
m_{\piz}$ between the \mtau lepton and the neutral pion. Using a
simplified analytical expression for the
form factor~\cite{Bergstroem1982}, the magnitude of $R$ is estimated to be around 0.2.
The SM electromagnetic \mtau-\piz coupling is therefore
\begin{equation}
  \left|g_{\tau\tau}^\text{e.m.}\right| \sim \order\left(10^{-5}\right),
\end{equation}
which is approximately four orders of magnitude smaller than the
coupling strength expected for the impostor model. 

A second potential SM background arises from events in which the \piz
meson is created through the $s$-channel virtual photon from the \epem
annihilation, together with another photon that converts to a
\mtau-lepton pair. This process is highly suppressed by the form
factor at $Q^2 = \left(10.58\gev\right)^2$. Compared to the
\mtau-lepton pair rate, it is further suppressed by a factor of $\alpha$.

The total combined expected background yield from the two
SM background processes described above corresponds to less 
than around 0.01 events, which is negligible compared to 
the number of pion impostor events required to explain the 
\fQQ  anomaly.

We present a search for new \piz-like particles in the
$\epem\to\tautau\impost$ final state, where \impost can be any of the
\impP, \impS, or \impHCP states. The paper is organized as follows:
Section~\ref{sec:babar} describes the detector and data samples used
in this analysis, while Section~\ref{sec:Analysis} presents the signal
selection and the yield extraction methodology.  The main
contributions to the systematic uncertainty are described in
Section~\ref{sec:Systematics} and the results are presented in
Section~\ref{sec:Results}. Section~\ref{sec:Summary} contains a
summary.

\section{\boldmath The \babar\ detector, data and simulation}
\label{sec:babar}
The data used in this analysis were collected with the \babar\
detector at the \pep2\ asymmetric-energy \epem\ storage rings between 1999 and
2007. The \babar\ detector is described in  detail elsewhere~\cite{Aubert2002,
  Aubert2013}. Here we provide a brief overview of the two
subdetectors most relevant to this analysis.

The energy of photons and electrons is measured with an electromagnetic
calorimeter (EMC) composed of a cylindrical array of CsI(Tl) crystals.
The resolution for the polar and azimuthal angles is
$\sim4\mrad$, and the energy resolution is
$\sim3\%$ for 1\gev photons~\cite{Aubert2002}.
The EMC also serves as a particle identification (PID) device for
electrons. The drift chamber is used to determine the momentum of the
charged tracks by measuring their curvature in a 1.5~T magnetic field.
The transverse momentum resolution is a linear function of the
transverse momentum \pt and is 0.67\% for $\pt=1.7\gevc$, which is
the mean laboratory \pt value of charged tracks expected in signal
events.

This analysis is based on 424\invfb of data collected at a
center-of-mass (CM)  energy \Eon\ and on 44\invfb collected at \Eoff~\cite{Lees2013}, 
corresponding to a total production of approximately $430\times10^6$ \tautau\
pairs.

\begin{figure}[ht]
  \centering
  \includegraphics[scale=1]{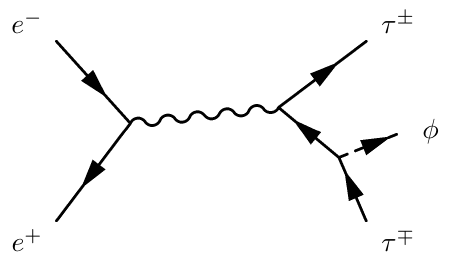}
  \caption{Diagram of the pion impostor production process in \epem
    annihilations. The \impost can be any of the \impP, \impS, or
    \impHCP particles.}
\label{fig:DiagEpemToTautauPi0}
\end{figure}

Simulated signal events are created using the {\tt  EvtGen}~\cite{Lange2001} generator.  First, large samples of
\epem\to \tautau\piz events are generated, based on three-body phase
space and nominal decay modes for the \mtau leptons and \piz meson.
Then the events are reweighted to reflect the production rate of
$\epem\to\tautau\impost$ processes using the analytical matrix
elements corresponding to the pion impostor process illustrated in
Fig.~\ref{fig:DiagEpemToTautauPi0}, assuming either the scalar or
pseudoscalar hypothesis.

The following backgrounds are considered: $\epem\to\BB$ events,
generated with the {\tt EvtGen}~\cite{Lange2001} program, continuum
hadronic $\epem\to\qqbar~(q=u,d,s,c)$ events, generated with
the {\tt JETSET}~\cite{Sjostrand1994} program, $\epem\to\mumu$ and
$\epem\to\tautau$ events, generated with the
$\mathcal{KK}$~\cite{Jadach1999} program, with the decay of the \mtau
leptons described using the {\tt TAUOLA}~\cite{Jadach:1993} library,
and \epem\to\epem events are simulated with the {\tt
  BHWIDE}~\cite{Jadach1997} program.  Radiative corrections are modeled with
the {\tt PHOTOS}~\cite{Barberio1994} algorithm and the detector response 
 with  the {\tt GEANT4}~\cite{Agostinelli2003} toolkit.


\section{\boldmath Analysis method}
\label{sec:Analysis}

The signal consists of a \tautau pair and a single pion impostor
\impost.  The pion impostor decays to a pair of photons with diphoton
invariant mass close to the \piz mass.
The selection criteria are optimized using simulated signal and
background events.  Simulated samples are also used to evaluate the
selection efficiency and systematic uncertainties.  These quantities
are evaluated using an impostor mass set equal to the mass of the
\piz.

\subsection{Signal selection}
\label{sec:SignalSelection}
For the selection of $\epem\to\tautau\impost$ signal events, we require
one $\tau$ lepton to decay leptonically to an electron and the other
to a muon.  This requirement suppresses background from radiative
Bhabha and dimuon events. We thus require events to contain exactly
two charged tracks, one identified as an electron and the other as a
muon. To reduce background from two-photon $\epem\to\epem X$ events,
signal event candidates are required to have a missing transverse
momentum larger than $0.3\gevc$, where the missing transverse momentum
is the magnitude of the vector sum of the \pt values of both tracks and
of all reconstructed neutral particles, evaluated in the event CM frame.

The pion-impostor candidates \impost are reconstructed by combining
two photons, each with a CM energy larger than 250\mev. To reduce the
contribution of radiative events, we require the sum of the CM energies
of all photons in the event not associated with the \impost candidate
to be less than 300\mev. The latter requirement also has the effect of
rejecting events containing more than one \impost candidate.
The photons associated with a \impost candidate must be separated from
the electron track by at least $30^\circ$ to further suppress
radiative events.  Control samples of
$\tau^\pm\to X^\pm \left(\piz\right) \nut$ events with $X^\pm = \pi^\pm$,
$K^\pm$, $\mu^\pm \num$ are used to determine
momentum-dependent corrections for the \impost selection
efficiency~\cite{Adametz2011}.

\begin{figure}[t]
  \centering
    \includegraphics[width=\columnwidth]{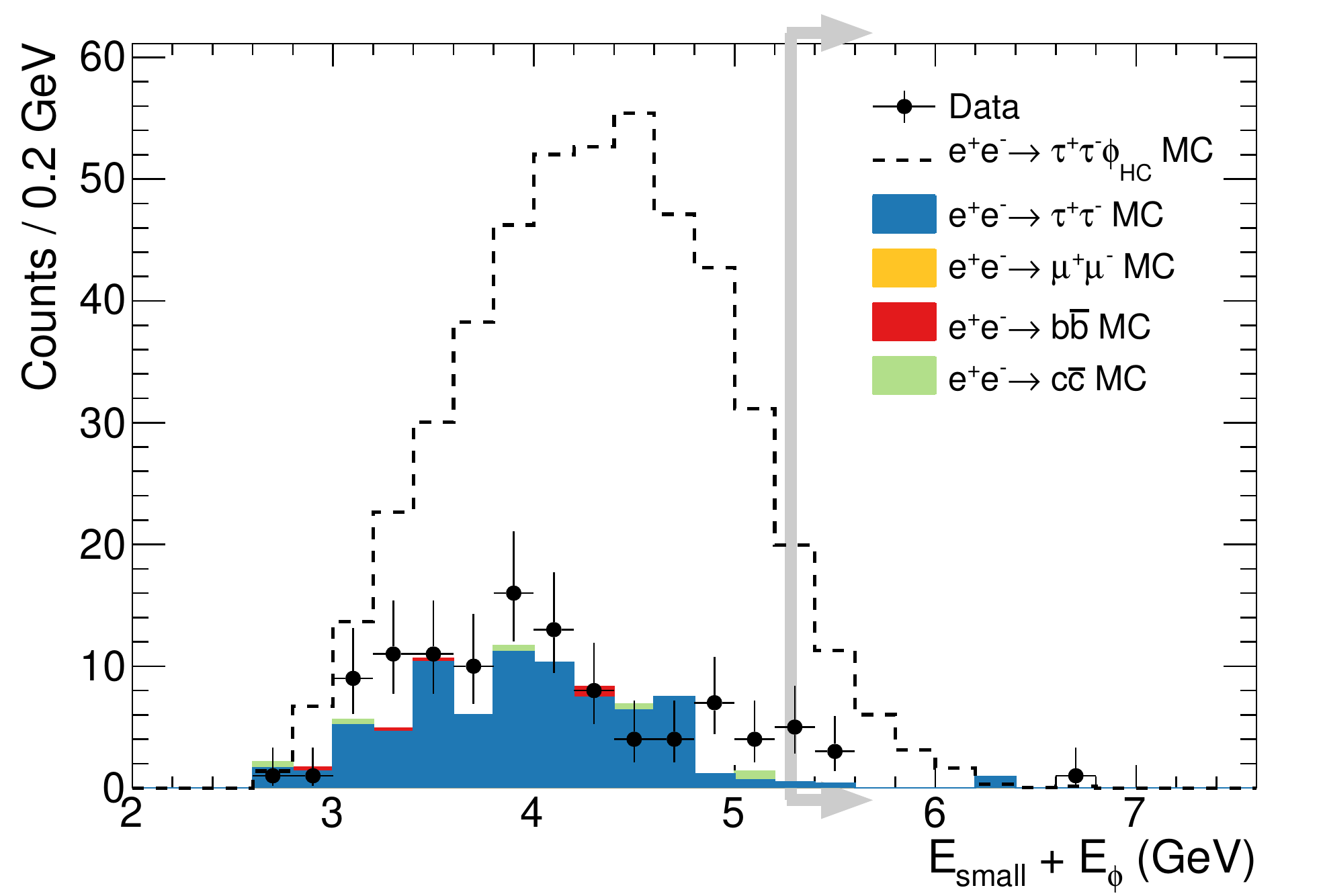}
    \includegraphics[width=\columnwidth]{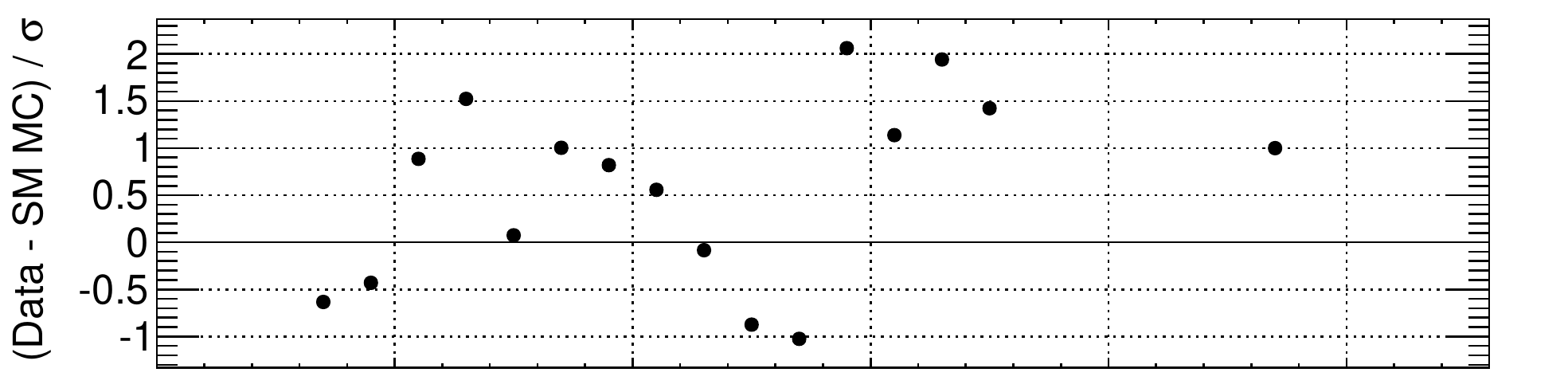}
  \caption{\coloronline Top: sum of the smaller of the track energies
    $E_{\rm small}$ and of the \impost candidate energy $E_{\impost}$,
    evaluated in the event CM, after applying all other selection
    criteria and requiring $m_{\gaga} \in \left[100,
      160\right]\mevcc$. The data to the right of the vertical line at
    5.29\gev are in the signal region. The predicted hardcore pion
     $\epem\to\tautau\impHCP$ distribution, assuming a
    production cross section of 0.254\pb, is included for reference.
    Bottom: Difference between data and Standard Model simulation (SM
    MC), divided by combined statistical uncertainty.}
  \label{fig:EsmEpi0}
\end{figure}

Kinematic constraints are used to ensure that the \impost candidate
does not arise from events in which one \mtau lepton decays
leptonically, while the other decays through $\tau^\pm \to
\rho^\pm\nu$ followed by $\rho^\pm\to\pi^\pm\piz$, where the $\pi^\pm$
is misidentified as a lepton.
 We form the invariant mass between  each track and the \impost candidate,
assuming a \pipm\ mass hypothesis for the track, and require the
combined mass to be greater than the \mtau-lepton mass. To further suppress
neutral pions from \mtau-lepton decays, the sum of the CM
energy of the \impost candidate, $E_{\impost}$, and that of the track
with the lower energy, $E_{\rm small}$, must be greater than
${\sqrt{s}}/{2}$.  The distribution of $E_{\rm small}+E_{\impost}$ for
events with $m_{\gaga} \in \left[100, 160\right]\mevcc$, after all
other selection criteria have been applied, is shown in
Fig.~\ref{fig:EsmEpi0}.

\begin{figure}[t]
  \centering
    \includegraphics[width=\columnwidth]{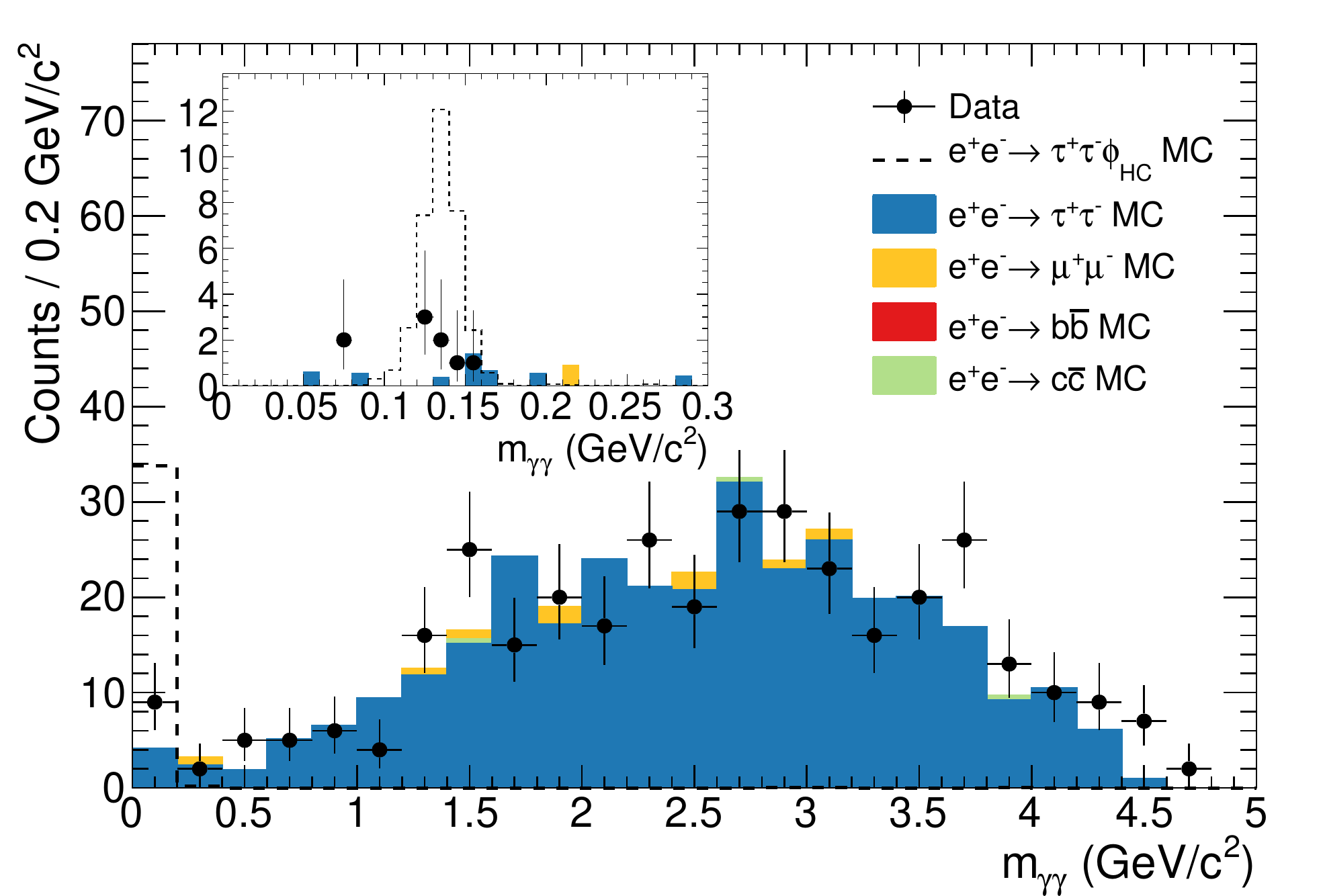}
  \caption{\coloronline Distribution of $m_{\gaga}$ after applying all
    other selection criteria. The insert shows the low
    mass range with bin size of 10\mevcc}
  \label{fig:Mpi0}
\end{figure}

The resulting diphoton mass spectrum after applying all other
selection criteria is displayed in Fig.~\ref{fig:Mpi0}. The data are
seen to agree with the SM simulation to within the uncertainties.

\subsection{Yield extraction, background evaluation, and selection efficiency}
\label{sec:BackgroundAndEfficiencies}
The signal yield is extracted by performing a series of extended
unbinned maximum likelihood
fits 
to the diphoton invariant mass distribution in the $\left[ 50,
  300\right]\mevcc$ range, scanning \impost mass hypotheses as
explained below. This region is chosen because it includes the
predicted mass range for the signal, and also because the background
distribution is relatively flat. The $m_{\gaga}$
distribution is fitted with the sum of a Gaussian function, describing
the contribution of the signal and peaking background components, and
a first-order polynomial representing the combinatorial background.
The number of events in the Gaussian peak is denoted $N_g$. The slope
and normalization of the polynomial as well as the value of $N_g$ are
determined in the fit. The mean $\mu_g$ and width $\sigma_g$ of the
Gaussian function are fixed to values determined as explained below.

The value of $\sigma_g$ is evaluated using control samples. These
samples are selected, for both data and simulation, using criteria
similar to those described above, but {reversing} the requirements on
the invariant mass formed from the charged track and the \piz
candidate, and {removing} the requirement on $E_\text{small} +
E_{\impost}$. The reason this latter requirement is removed is to
increase the statistical precision. 
The $m_{\gaga}$ spectra are then fitted using the signal model
described above except with $\sigma_g$ a fitted parameter.
We find $\sigma_g = 10.6 \pm 1.8 \mevcc$ for the data and $\sigma_g =
11.2 \pm 0.8 \mevcc$ for the simulation.  For the subsequent fits,
we fix $\sigma_g$ to $11.1 \mevcc$, which is the average of the results
from data and simulation. 

The value of $\mu_g$ represents the mass of the hypothetical \impost
particle. It is is fixed in the fit and scanned between 110 and
160\mevcc, covering the expected range of impostor mass
values~\cite{McKeen2012}.  The step size is 0.5\mevcc, corresponding
to less than half the estimated mass resolution.

We select the scan point that yields the largest value $N_g^{\rm max}$
of $N_g$.  The signal yield $N_{\rm sig}$ is obtained by subtracting
the estimated number of peaking background events from $N_g^{\rm max}$
and correcting for the signal yield bias.

The number of peaking background events predicted by the simulation is
$0.38\pm0.09$, where the uncertainty accounts for uncertainties in the
PID as well as for the difference between the data and simulation rates in
the sidebands, which is visible in Fig.~\ref{fig:EsmEpi0} for values
of $E_{\rm small} + E_\impost$ above 4.8\gevcc.

We also consider potential peaking backgrounds that are not present in
the simulation.  Specifically, we consider two-photon
$\epem\to\epem\pipi\piz$ events, for which either the $e^+$ or $e^-$,
and one of the charged pions, are undetected, while the other charged
pion is misidentified as a muon.  The events are selected using the
same criteria as described above except requiring the presence of a
charged pion rather than a muon.  The $m_{\gaga}$ spectrum of the
selected events is fitted as described above, and the resulting value
of $N_g$ is scaled by the muon-to-pion misidentification rate of
$(3.0\pm1.0)\%$.  Adding the resulting value to the
number of peaking events determined from simulation yields an estimate
of $1.24\pm 0.37$ events. This number is subtracted from $N_{g}^{\rm
  max}$ as described above.

The evaluation of the fit bias is performed using a large ensemble of
pseudo-experiments.  For this purpose, diphoton invariant mass spectra
are generated to reproduce the combinatorial background with the
number of combinatorial events drawn from a Poisson distribution whose
mean equals the simulated result.  A peaking component centered at the
\piz mass is added. The number of peaking events is drawn from a
Poisson distribution with mean equal to one. Each peaking background
event is then weighted by a number drawn from a Gaussian distribution
whose mean and width are 1.24 and 0.37 events, respectively.  We
determine the bias for several values of the signal yield by further
adding a known number of signal-like events to each experiment.
Between 0 and 25 signal events are added to each pseudo-experiment,
yielding an average fit bias of $-0.06\pm0.02$~events.

The signal selection efficiency is determined by applying the analysis
procedures to the simulated signal events. After accounting for the
$\taum\to\mun\numb\nut$ and the $\taum\to \en \nueb \nut$ branching
fractions~\cite{Beringer2012}, the efficiencies are found to be
$\eps_{\impP}=\eps_{\impHCP} = (0.455\pm0.017)\%$ and $\eps_{\impS} =
(0.0896\pm0.0033)\%$, where the uncertainties are statistical.  The
efficiency to reconstruct the \impS is smaller than that to
reconstruct the \impP and \impHCP because the scalar particle tends
to produce lower-energy impostor candidates that do not satisfy the
selection criteria.


\section{\boldmath Systematic uncertainties}
\label{sec:Systematics}
Sources of systematic uncertainty affecting the efficiency measurement
include those associated with the \piz\ and PID efficiency
corrections, as well as differences between the data and simulation in the
track momentum scale and resolution, and in the photon energy scale
and resolution. These multiplicative uncertainties are summarized in
Table~\ref{tab:Systematics}. The additive uncertainty contributions to
the signal yield measurement are associated with the peaking
background estimate and potential biases in the fit procedure.
For the latter, we assign the full bias correction as a systematic
uncertainty.

The uncertainty related to the \piz reconstruction efficiency
 is evaluated by performing the analysis while varying the
\piz efficiency correction within its uncertainties. The PID
uncertainty is $0.5\%$, estimated using high-purity control samples.

The uncertainties associated with the differences between the data and
simulation for the track momentum scale and resolution are measured
using $\epem\to\mumu\gamma$ events. These samples are also used to
determine the uncertainties related to the photon energy scale and
resolution~\cite{Aubert:2009qj}.

\begin{table}[ht!]
  \centering
  \caption{Contributions to the uncertainty of the efficiency (\%) for the three models considered.}
  \begin{ruledtabular} 
 \begin{tabular}{ldd}
  Source of uncertainty & \multicolumn{1}{c}{$\impP,\impHCP$}
                        & \multicolumn{1}{c}{$\impS$}  \\
                        & \multicolumn{1}{c}{$(\%)$}
                        & \multicolumn{1}{c}{$(\%)$}  \\

  \colrule
  \quad MC sample size        & 3.5        & 3.7 \\
  \quad \piz\ efficiency      & 1.0        & 1.0 \\
  \quad PID                   & 0.5        & 0.5 \\
  \quad Momentum scale        & 0.2        & 0.2 \\
  \quad Momentum resolution   & 0.1        & <0.1 \\
  \quad Energy scale          & 2.0        & 2.0 \\
  \quad Energy resolution     & 0.6        & 0.6 \\
  \colrule
  Total systematic uncertainty
                              & 4.2        & 4.4\\
  \end{tabular}
  \end{ruledtabular}
  \label{tab:Systematics}
\end{table}


\section{\boldmath Results}
\label{sec:Results}

\subsection{Data $\textit{m}_{\gaga}$ spectrum}

\begin{figure}[t]
  \centering
  \includegraphics[width=\columnwidth]{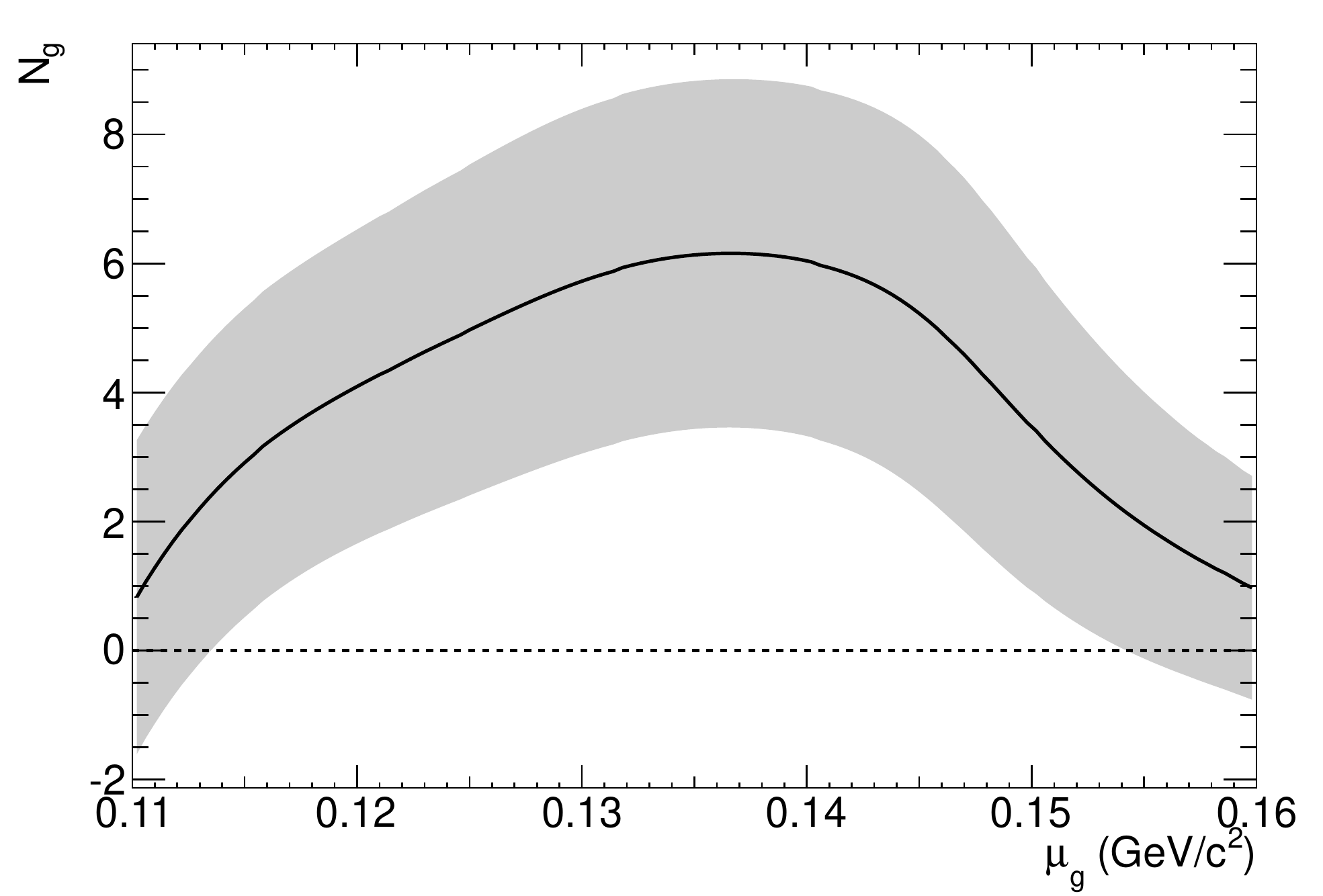}
  \caption {Number $N_g$ of events in the Gaussian peak as a function of the
    \impost mass hypothesis $\mu_g$.  The shaded region indicates the statistical
    uncertainty.}
  \label{fig:Mgaga_Scan}
\end{figure}

\begin{figure}[t]
  \centering
  \includegraphics[width=\columnwidth]{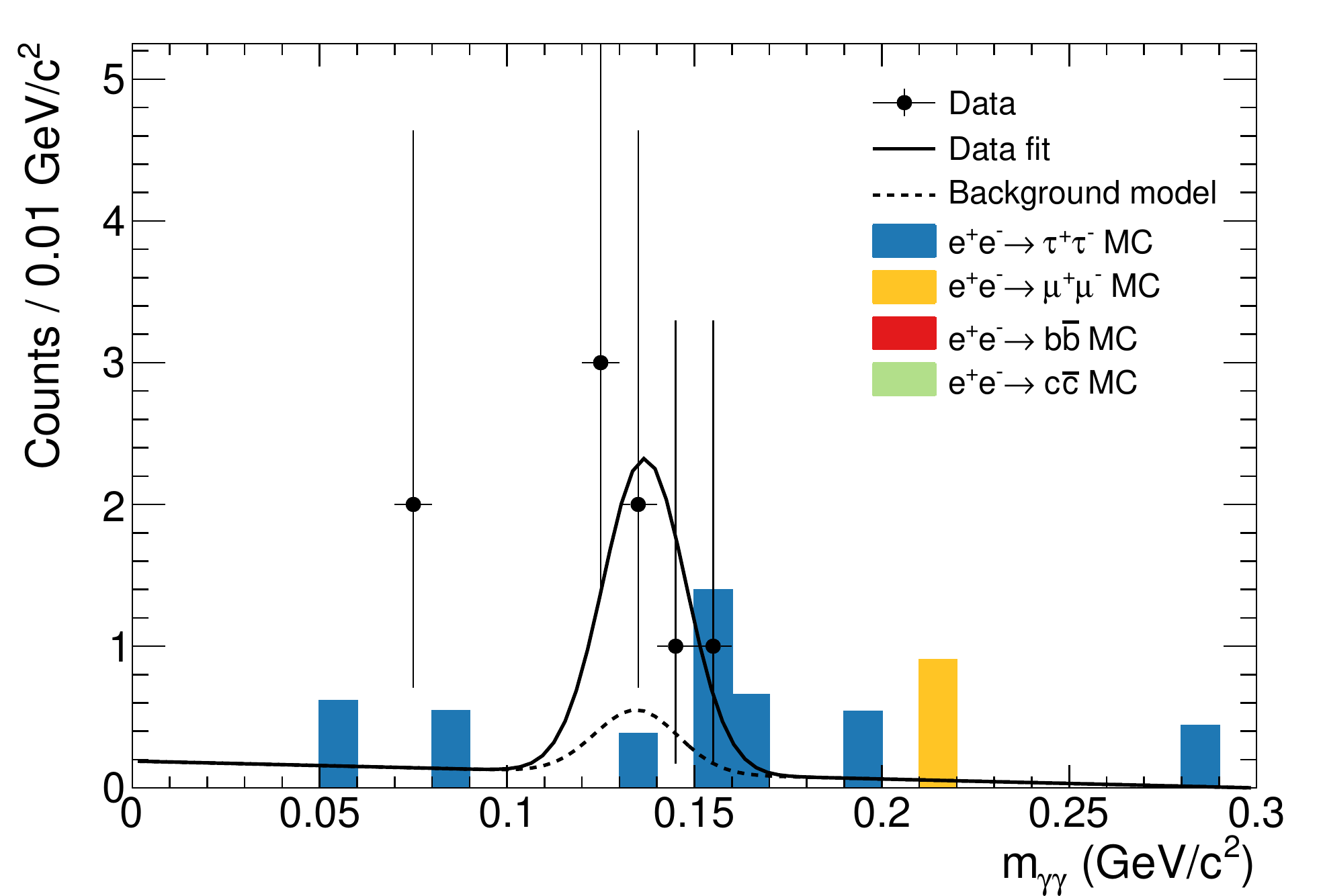}
  \caption{\coloronline Results for the $m_{\gaga}$ spectrum of the
    signal candidates. The solid line shows fit result for the signal
    and background model. The dotted line represents the
    contribution from background only using the linear component of the
    fit result added to the estimated peaking background of 1.24~events.}
  \label{fig:DataMgaga}
\end{figure}

Figure~\ref{fig:Mgaga_Scan} shows the yield $N_g$ of events in the
Gaussian peak, with its statistical uncertainty, as a function of the
\impost particle mass hypothesis.
The largest value, $N_{g}^{\rm max}=6.2\pm2.7\stat$~events, arises for $\mu_g =
136\mevcc$. The fit result with this mass hypothesis is shown in the diphoton
mass distribution of Fig.~\ref{fig:DataMgaga}, where the
contribution from the expected background is also presented.  The
probability of observing a signal of at least 6.2 events assuming a
background-only hypothesis is estimated from the pseudo-experiments
described in Section~\ref{sec:Systematics}, which assume a mass
$\mu_{g} \in \left[110, 160\right]\mevcc$.  The $p$-value is found to
be $p_0 = 3.71\times10^{-2}$.

After subtraction of the peaking background and correction for the fit
bias, the number of signal candidate events at $\mu_g=136\mevcc$ is found to be
\begin{equation}
N_{\rm sig} = 5.0\pm2.7\stat\pm0.4\syst.
\end{equation}
Correcting this result for the signal selection efficiency leads to
the following production cross sections:
\begin{equation}
\sigma_{} =
\left\{
\begin{array}{rl}
38 \pm 21 \stat \pm 3\syst\fb & \text{ for \impP  and  \impHCP,}\\
190  \pm 100  \stat \pm 20 \syst\fb & \text{ for \impS.}
\end{array}
\right.
\end{equation}
Statistical uncertainties dominate in both cases. The main source of
systematic uncertainty is the peaking background estimation and
subtraction procedure.

\subsection{Upper limits on the cross sections}
No significant signal is observed. Upper limits on the production
cross sections are set using the $CL_s$ method~\cite{Read2002}. The
90\% confidence level (CL) upper limit on the number of signal events,
$N_{\rm sig} \leq 9.6$, translates into the following bounds on the
cross section
\begin{equation}
\sigma_{} \leq
\left\{
\begin{array}{rl}
73\fb & \text{ for the \impP and \impHCP models,}\\
370\fb & \text{ for the \impS model.}\\
\end{array}
\right.
\end{equation}

\subsection{Compatibility of the measurement with the \piz\ impostor theories}
The compatibility of the measured production cross sections with the
impostor theories is studied by including this measurement as an
additional term in the $\chi^2$ when calculating the optimal coupling
values needed to describe the \babar\ measurement of \fQQ. The
increase in $\chi^2$ obtained when adding the couplings corresponding
to our cross section measurements follows a $\chi^2$ distribution with
one degree of freedom.  This is used to determine the $p$-values
corresponding to a fluctuation of the $\epem\to\tautau\impost$ event
rate from the level seen in the present study to the level required to
explain the \babar{} \fQQ measurements.

The results are reported in Table~\ref{tab:CompChi2}.  As an example,
the $p$-value for the hardcore pion model is found to be $5.9\times
10^{-4}$, corresponding to a required fluctuation of 3.4 standard
deviations.  The $p$-values for the \impP and \impS models are on the
order of $10^{-9}$.  Thus the pion impostor models do not provide a
likely explanation for the excess seen in the \babar{} pion-photon
transition form factor data.
\begin{table}[ht]
  \centering
  \caption{Consistency ($p$-value) of the measured production
    cross sections with the impostor theories adjusted to the \babar\ \fQQ data.}
  \begin{ruledtabular}
  \begin{tabular}{ccccl}
    Model & \multicolumn{2}{c}{$\chi^2_\text{min}/\text{n.d.f.}$}
          &  $\Delta \chi^2 / \text{n.d.f.}$ & \multicolumn{1}{c}{$p$-value} \rule{0pt}{1.2eM}\\
          & \fQQ and  & \fQQ  \rule{0pt}{1.2eM}\\
          &$\sigma_{\epem\to\tautau\impost}$ & only\\
   \colrule
   \impHCP & 23.7/10 & 11.9/9 & 11.8/1 & $5.9 \times 10^{-4}$\rule{0pt}{1.2eM}\\
   \impP   & 48.4/10 & 10.8/9 & 37.6/1 & $8.8 \times 10^{-10}$\rule{0pt}{1.2eM}\\
   \impS   & 49.2/10 & 13.4/9 & 35.8/1 & $2.2 \times 10^{-9}$\rule{0pt}{1.2eM}\\
  \end{tabular}
\end{ruledtabular}
\label{tab:CompChi2}
\end{table}


\section{Summary} \label{sec:Summary}

A search for \piz\ impostors  is conducted with the
\babar\ data set. At 90\% confidence level, the limit on the
production cross section in association with a \tautau pair is 73\fb
for the pseudoscalar impostor and the hardcore pion models, and
370\fb for the scalar impostor model. The $p$-values of our measurements
under these hypotheses are $5.9\times10^{-4}$ or smaller. The pion
impostor hypotheses are disfavored as explanations for the
non-asymptotic behavior of the pion-photon transition form factor
 observed with the \babar\ data.
\begin{acknowledgments}
  \input{acknow_PRL}
\end{acknowledgments}

\bibliography{references}

\end{document}

%% file: authors_aug2014_bad2598.tex
%
\author{J.~P.~Lees}
\author{V.~Poireau}
\author{V.~Tisserand}
\affiliation{Laboratoire d'Annecy-le-Vieux de Physique des Particules (LAPP), Universit\'e de Savoie, CNRS/IN2P3,  F-74941 Annecy-Le-Vieux, France}
\author{E.~Grauges}
\affiliation{Universitat de Barcelona, Facultat de Fisica, Departament ECM, E-08028 Barcelona, Spain }
\author{A.~Palano$^{ab}$ }
\affiliation{INFN Sezione di Bari$^{a}$; Dipartimento di Fisica, Universit\`a di Bari$^{b}$, I-70126 Bari, Italy }
\author{G.~Eigen}
\author{B.~Stugu}
\affiliation{University of Bergen, Institute of Physics, N-5007 Bergen, Norway }
\author{D.~N.~Brown}
\author{L.~T.~Kerth}
\author{Yu.~G.~Kolomensky}
\author{M.~J.~Lee}
\author{G.~Lynch}
\affiliation{Lawrence Berkeley National Laboratory and University of California, Berkeley, California 94720, USA }
% ...
\collaboration{The \babar\ Collaboration}
\noaffiliation

%% file: acknow_PRL.tex
We are grateful for the excellent luminosity and machine conditions
provided by our \pep2\ colleagues, 
and for the substantial dedicated effort from
the computing organizations that support \babar.
The collaborating institutions wish to thank 
SLAC for its support and kind hospitality. 
This work is supported by
DOE
and NSF (USA),
NSERC (Canada),
CEA and
CNRS-IN2P3
(France),
BMBF and DFG
(Germany),
INFN (Italy),
FOM (The Netherlands),
NFR (Norway),
MES (Russia),
MINECO (Spain),
STFC (United Kingdom),
BSF (USA-Israel). 
Individuals have received support from the
Marie Curie EIF (European Union)
and the A.~P.~Sloan Foundation (USA).
%
%
%